\newcommand {\be}{\begin{equation}}
\newcommand {\ee}{\end{equation}}

\newcommand {\bea}{\begin{eqnarray}}
\newcommand {\eea}{\end{eqnarray}}

\documentstyle[twocolumn,aps,prl]{revtex}
\draft

\begin{document}

\twocolumn[\hsize\textwidth\columnwidth\hsize\csname @twocolumnfalse\endcsname

\title{
Quantum-critical scaling and temperature-dependent logarithmic corrections
in the spin-half Heisenberg chain}
\author{O. A. Starykh$^{1}$, 
 R. R. P. Singh$^{1}$, and A. W. Sandvik$^{2}$\cite{byline1}}
\address{$^{1}$Department of Physics, University of California, Davis, 
California 95616\\
$^{2}$National High Magnetic Field Laboratory, 1800 East Paul Dirac Drive,
Florida State University, Tallahasse, FL 32306}
\date{\today}
\maketitle{}

\begin{abstract}
Low temperature dynamics of the $S={1\over 2}$ Heisenberg chain is studied
via a simple ansatz generalizing the conformal mapping and
analytic continuation procedures to correlation functions with multiplicative
logarithmic factors. Closed form expressions for the dynamic
susceptibility and the NMR relaxation rates $1/T_1$ and $1/T_{2G}$ are 
obtained, and are argued to improve the agreement with recent experiments.
Scaling in $q/T$ and $\omega/T$ are violated due to these
logarithmic terms. Numerical results
show that the logarithmic corrections are very robust.
While not yet in the
asymptotic low temperature regime, they provide striking 
qualitative confirmation of the theoretical results.
\end{abstract}
\vskip2mm]

In recent years there has been much interest in
quantum critical phenomena in spin-models and real materials.
In path integral formulations, the inverse temperature $\beta$
acts as a finite size for the imaginary-time fluctuations,
driving the system away from the $T=0$ quantum critical point. 
The resulting behavior can be described in a manner analogous
to finite-size scaling. In two-dimensions, quantum critical points
are rare, but their relevance to real materials is enhanced by the fact that
they also control the finite temperature properties
of weakly-ordered or weakly-gapped systems \cite{CHNCSY}. 

In contrast to $2D$, $1D$ quantum antiferromagnets
with continuous symmetry are generically critical at $T=0$.
Thus one expects to find many examples of quantum critical phenomena
in quasi-one dimensional materials. Recently, quantum critical
behavior of spin-chains have been studied by neutron scattering \cite{Neutron}
and NMR experiments \cite{Takigawa}. 

From a theoretical point of view, the development of conformal field
theory provides a powerful machinery to study the finite temperature
correlation functions at a $T=0$ critical point.
Assuming a simple power-law
behavior for the $T=0$ correlation functions, various authors
have obtained scaling forms for the dynamic structure factor
at low temperatures \cite{Schulz,Shankar,SSS}.
However, it is well known that the $T=0$ spin-spin correlations of
the $S=1/2$ chain, have multiplicative logarithmic factors
due to the presence of marginally irrelevant operators \cite{Giam}:
\bea
<S(0,0)S(x,t)>_{T=0} = &&(-1)^x \frac{D}{\sqrt{x^2 - (ct)^2}}
\nonumber\\
&&\times
\left(\ln\frac{\sqrt{x^2 - (ct)^2}}{r_0}\right)^{1/2}.
\label{1}
\eea
Here $c$ is the spin wave velocity, and $D$ and $r_0$ are
nonuniversal constants.
These logarithmic factors also appear in the two-spinon
contribution to the dynamic structure factor \cite{2-spinon}
$S_{zz}^{(2)}(q,\omega) \sim \frac{1}{\sqrt{\omega^2 - (cq)^2}}
\sqrt{\ln{\frac{1}{\omega^2 - (cq)^2}}}$.

In this paper, we explore the effects of logarithmic factors on
the finite temperature dynamics
of the spin-half Heisenberg chain. We begin by considering the $T=0$ equal-time
correlations of a finite system. It was proposed in
Ref.~\cite{KM} that, in presence of logarithms, the correlations for
a system of size $L$ should have a generalized finite-size scaling form:
\bea
<S(0)S(x)>_L = &&(-1)^x D \left[\frac{1}{LX(x/L)}\right]^{d-2+\eta}
\nonumber\\
&&\times
\left(\ln(\frac{LX(x/L)}{r_0})\right)^{\sigma}.
\label{2}
\eea
Here the universal function $X(x/L)$ is
given by $X(x)=\frac{1}{\pi} \sin(\pi x)$ \cite{Cardy}.
This proposition was shown
to work well for the isotropic Heisenberg chain
with $\sigma=1/2$ \cite{KM}.

In the absence of logarithmic factors, correlations at finite $T$ are 
obtained from a conformal mapping of the complex plane $z=x + ic\tau$ at 
$T=0$ to a strip infinite in the $x$-direction and of width $c/T$ in the 
$\tau$-direction ($\tau$ is imaginary time). In close analogy with $T=0$ 
finite-size scaling, this amounts to the substitution
\be
LX(\frac{z}{L},\frac{\bar{z}}{L}) \rightarrow \frac{c}{\pi T}
\left(\sinh(\pi T(\frac{x}{c} + i\tau)) \sinh(\pi T(\frac{x}{c} - i\tau))
\right)^{1/2}.
\label{3}
\ee
We now assume that this mapping can also be performed in presence
of marginal interactions. Then finite $T$ spin correlations
are given by the ansatz (\ref{2}) with scaling function (\ref{3}).
This can be further approximated as:
\bea
&&<S(0,0)S(x,\tau)>_T = (-1)^{x} D \frac{\sqrt{2} \pi T}{c} 
\left(\ln{\frac{T_0}{T}}\right)^{1/2}
\nonumber\\
&&\times
\left(\cosh{\frac{2\pi T x}{c}} - \cos{2\pi T \tau}\right)^{-2\Delta},
\label{4}
\eea
where $T_0 = \sqrt{2}\pi c/{r_0}$, and an
effective $temperature-dependent$ scaling dimension appears
\cite{AB,KM}
\be
\Delta=\frac{1}{4} \left(1 - \frac{1}{2\ln{\frac{T_0}{T}}}
\right).
\ee
Equation (\ref{4}) is valid for $x \ll \xi \ln\frac{T_0}{T}$ (see
expression for $\xi$ below), and thus allows one to study correlations
below and above the correlation length.
An immediate consequence of this ansatz is that 
the correlation length aquires a logarithmic temperature
dependence, in agreement with thermal Bethe ansatz calculations 
\cite{Nomura}:
\be
\xi^{-1} = \frac{\pi T}{c} (1 - \frac{1}{2\ln{\frac{T_0}{T}}}).
\label{5}
\ee

We now explore further consequences of the scaling ansatz. First,
the static structure factor is found to be
\bea
S(q)= &&2^{2\Delta + 1/2} D \left(\ln{\frac{T_0}{T}}
\right)^{1/2}
\Gamma(1 - 4\Delta) 
\nonumber\\
&&\times
Re\left( \frac{\Gamma(2\Delta - i\frac{cq}{2\pi T})}{\Gamma(1 - 2\Delta -
i\frac{cq}{2\pi T})} 
\right),
\label{7}
\eea
where $q$ is measured from the antiferromagnetic vector $\pi$.
Note that the entire q-dependence of $S(q)$ is due to the $1/\ln{T}$ 
corrections to the $T=0$ value of $\Delta = 1/4$.
Eq.~(\ref{7}) implies that $S(q)/S(0)$ is no longer a universal function
of $cq/T$.

Performing Fourier transformation and analytic continuation to real
frequencies \cite{Schulz,Shankar,SSS}, we obtain the staggered
susceptibility
\bea
&&\chi(q,\omega)= \frac{2^{2\Delta - 3/2} D}{\pi T}\sin(2\pi\Delta)
\left( \ln{\frac{T_0}{T}}\right)^{1/2} 
\Gamma^2(1 - 2\Delta)
\nonumber\\
&&\times
\frac{\Gamma(\Delta - i\frac{\omega - cq}{4\pi T})}{\Gamma(1 - \Delta
- i\frac{\omega - cq}{4\pi T})} \frac{\Gamma(\Delta - i\frac{\omega + cq}
{4\pi T})}{\Gamma(1 - \Delta- i\frac{\omega + cq}{4\pi T})}.
\label{8}
\eea
This expression also lacks universality due to the T-dependence of $\Delta$. 

Next, we discuss numerical results for the spin-half chain obtained using
a ``stochastic series expansion'' Quantum Monte Carlo (QMC) method 
\cite{Sandvik1} (for systems with up to 1024 spins), and conventional 
high temperature expansions (HTE). Most results from the 
two methods agree down 
to $T/J=1/8$. Below that temperature, we rely on QMC data alone.

We begin with the $\omega=0$ susceptibility, shown in Fig.~\ref{chi}.
The ratio $\chi(q,0)/\chi(0,0)$ appears to converge towards a scaling 
form as the temperature is lowered, but even at $\beta=32$ it is far
from the universal scaling function expected in the absence of logarithms 
\cite{Sachdev}. In the range $1/4>T>1/8$ the numerical results have high 
accuracy, and QMC and HTE data agree very well. The 
deviations from scaling are clearly systematic, and well described by 
Eq.~(\ref{8}) with $T_0=4.5$. Note that the parameter $T_0$ should be 
considered an {\it effective} one. As the study of logarithmic corrections 
to the uniform susceptibility shows \cite{Eggert}, the {\it true} value of 
$T_0$ maybe reached only at $T\le 0.01$.

Data for $S(q)$ show substantial $q$-dependence, in disagreement 
with $\Delta=1/4$ scaling predictions. However, the results are not well 
explained by Eq.~(\ref{7}) either. A possible reason is that $S(q)$ is 
dominated by contributions (divergent at $T=0$) from short distances, 
where our asymptotic expression (\ref{4}) breaks down. It is, thus, better 
to compare the equal-time real-space spin correlations, $S(x)$, with the 
theoretical expressions. It is well known that the correlation function
in addition to the dominant staggered piece has a uniform contribution, 
given by  $-({T\over 2c\sinh{(\pi Tx/c)}})^2$ at finite $T$ \cite{Sachdev}. 
It is appropriate to subtract this from the numerical data 
before comparing with the scaling theory. As shown in Fig.~\ref{Sr},
our results for $S(x)$ agree very well with Eq.~(\ref{4}),
with $T_0=4.5$ and $D=0.075$. The inset shows a comparison of the
ratio of correlation functions at two temperatures. With $T_0$ fixed
from the susceptibility data, this parameter-free agreement is
quite striking. Deviation of the theoretical results at short distances
is also apparent and is the reason that $S(q)$ cannot be
explained. The theoretical results also imply $S(0)\sim (\ln{\beta})^{3/2}$
and $\chi(0,0) \sim \beta(\ln{\beta})^{1/2}$ as $T\to 0$,
in agreement with numerical data \cite{Starykh}.

From Eq.~(\ref{8}) we can calculate the NMR relaxation rates
\begin{equation}
\frac{1}{T_1} =\frac{2^{5/2 - 2\Delta} A_{\parallel}^2(\pi) D}
{\pi c}\sin(2\pi\Delta) I_1(\Delta)
\left(\ln{\frac{T_0}{T}}\right)^{1/2},
\label{9}
\end{equation}
\bea
\frac{1}{T_{2G}} =&&\frac{2^{-3 + 2\Delta} A_{\perp}^2 (\pi) D}
{\pi c} \sin(2\pi\Delta) \Gamma^2(1 - 2\Delta) I_2(\Delta)
\nonumber
\\
&&\times
\sqrt{\frac{c}{T}\ln{\frac{T_0}{T}}}.
\label{10}
\eea
Here the integrals $I_1(\Delta)=\int_0^{\infty} dx \frac{x}
{(\sinh{x})^{4\Delta}}$ and $I_2^2(\Delta)=4\int_0^{\infty}
|\frac{\Gamma(\Delta - ix)}{\Gamma(1 - \Delta - x)}|^4$ have
weak temperature dependences. 
In deriving Eq.~(\ref{10}), we have kept only the scaling part
and dropped the term coming from
self interactions as it is down by a factor
$T(\ln{\frac{T_0}{T}})^2$. 
The latter is, in any case, not correctly accounted for
by the scaling theory.
We find that Eq.~(\ref{9}) shows weaker than $\sqrt{\ln{T_0/T}}$
variations with $T$. This result is in qualitative agreement with 
recent measurement of $1/T_1$ in Sr$_{\rm 2}$CuO$_{\rm 3}$
by Takigawa {\it et al.} \cite{Takigawa}. Fig.~\ref{ratio} shows the ratio 
$T_{2G}/\sqrt{T}T_1$. In the $T \rightarrow 0$ limit our expressions 
(\ref{9}) and (\ref{10}) coincide with those of Sachdev \cite{Sachdev}.
However, we find that $T=0$ limit of $T_{2G}/\sqrt{T}T_1$ is approached
with infinite slope, similar to the behavior of the uniform susceptibility
\cite{Eggert}. The behavior is consistent with the slow rise of this
quantity seen for Sr$_{\rm 2}$CuO$_{\rm 3}$ around $T=J/10$ \cite{Takigawa}. 

Numerical results for $\chi^{\prime\prime}$ are obtained from QMC data using 
the maximum-entropy (max-ent) method \cite{maxent}, and from HTE
via the recursion method \cite{Starykh}. In Fig. \ref{ratio}, 
data are presented for the ratio with the full $T_{2G}$ and with only the 
scaling part, where the self-term is not subtracted \cite{footnote}. The 
two should converge in the scaling limit and the {\it latter} should be 
compared with the theoretical result. Note that QMC and HTE
results agree completely for $T_{2G}$; the deviations arise
entirely from the analytic continuation needed to get $1/T_1$, which is
more uncertain for QMC data (details of this point will be discussed 
elsewhere \cite{Starykh}). The difference between the curves based on
the full $T_{2G}$ and the scaling part only of $T_{2G}$ shows that the 
results are not yet in the scaling limit. However, the theoretical results 
are supported by the convergence of the more accurate (in the temperature 
regime shown) HTE data to the predicted form. The presence of non-asymptotic 
contributions in the full $T_{2G}$ and the apparent tendency of QMC + 
max-ent to over-estimate $1/T_1$ explain the discrepancy in the numerical
result previously reported for $T_{2G}/\sqrt{T}T_1$ \cite{Sandvik}. 

We also note that the experimental $1/T_1$ was found to be about $30$\% 
lower than the numerical result at $T=300$K \cite{Takigawa,Sandvik}, which
is now also reconciled. Together with the good agreement found previously for 
$1/T_2$, without adjustable parameters \cite{Takigawa,Sandvik}, the spin-half 
chain indeed very well describes the low-frequency dynamics of 
Sr$_{\rm 2}$CuO$_{\rm 3}$. 

The frequency-dependent
quantities also do not show universality in the
scaled variable $\omega/T$.
For example,
the imaginary part of the $q=0$ and local 
($\chi(\omega)=\int_{-\infty}^{\infty}\frac{d q}{2\pi} \chi(q,\omega)$)
susceptibility are given,
respectively, by
\bea
&&\chi^{\prime \prime}(0,\omega)=\frac{2^{2\Delta - 3/2} D}{\pi T} 
\sin(2\pi\Delta)
\left(\ln{\frac{T_0}{\sqrt{2}\pi T}}\right)^{1/2} \Gamma^2(1-2\Delta),
\nonumber\\
&&\times
Re\left(\frac{\Gamma(\Delta + i\frac{\omega}{4\pi T})}{\Gamma(1 - \Delta
+ i\frac{\omega}{4\pi T})}\right)
Im \left(\frac{\Gamma(\Delta + i\frac{\omega}{4\pi T})}
{\Gamma(1 - \Delta + i\frac{\omega}{4\pi T})}\right).
\label{11}
\eea
\bea
\chi^{\prime \prime}(\omega)=&&
\frac{2^{2\Delta - 1/2} D}{c} \sin(2\pi\Delta)
\left(\ln{\frac{T_0}{T}}\right)^{1/2} 
\Gamma(1 - 4\Delta)
\nonumber\\&&\times
Im \left(\frac{\Gamma(2\Delta - i\frac{\omega}{2\pi T})}
{\Gamma(1 - 2\Delta - i\frac{\omega}{2\pi T})}\right).
\label{12}
\eea
The $\omega$-dependence in Eq.~(\ref{12}) is due to the T-dependence
of the scaling dimension. The  $\omega$-independence and divergence at 
$\Delta=1/4$ is an artifact of the approximations employed, and should 
be removed by more accurate treatment of the short-distance cut-off. 
Nevertheless, fixing the scaling dimension at $1/4$ implies a much weaker
$T-$ and $\omega-$ dependence than predicted by (\ref{12}).
Note also that we predict 
$\chi^{\prime \prime}(0,\omega) \sim (\ln\frac{T_0}{T})^{1/2}$ and
$\chi^{\prime \prime}(\omega) \sim (\ln\frac{T_0}{T})^{3/2}$ 
at low temperatures.

QMC + max-ent data for the $q=0$ susceptibility, as well
as the results of Eq.~(\ref{11}), are shown in Fig.~\ref{chi''}, where the 
value of $T_0$ is from the fit of the static susceptibility.
At low frequencies the theory and the data agree and also
appear to scale. At higher frequencies, there is no
scaling and the deviations are qualitatively similar in that the
lower temperature data is higher at higher values of $\omega/T$.
The numerical data are not at low enough temperatures to
explore the scaling forms at larger $\omega/T$.
There are preliminary reports of measurements
of these quantities by neutron
scattering \cite{Neutron2}. 
It would be useful to compare them with our results.

To conclude, we have studied the effects of logarithmic corrections
on the finite temperature dynamic spin-correlations
of the spin-half chain. Analytical
expressions are developed for $\chi(q,\omega)$
by a generalized finite-size scaling ansatz. The ansatz ties together 
previous results, including logarithmic corrections to the correlation 
length, and implies a temperature dependent effective scaling dimension.
Expressions obtained for the NMR relaxation rates are argued to 
improve the agreement with experimental data for 
Sr$_{\rm 2}$CuO$_{\rm 3}$ \cite{Takigawa}. Numerical
results, although not in the asymptotic low temperature regime,
confirm various theoretical expressions including violation of scaling 
in the variables $cq/T$ and $\omega/T$. We expect these effects to diminish
and scaling to be restored if the second-neighbor interactions 
are tuned to the point
where the marginal interaction is absent \cite{Eggert2}.
We hope our calculations will provide further motivation for
neutron scattering studies of quasi-one dimensional 
spin systems.

Support from the NSF through grant numbers
DMR-9318537 (O.A.S and R.R.P.S) 
and DMR-9520776 (A.W.S) is gratefully acknowledged.

\begin{figure}
\caption
{The static susceptibility normalized to its $q=0$ value.  
Symbols represent numerical data from high-temperature expansions (HTE)
and Monte Carlo simulations (MC).
Solid lines are
predictions of Eq.~(\protect\ref{8})
with $T_0=4.5$, and the dashed line
shows the universal scaling function 
with $\Delta=1/4$
\protect\cite{Sachdev}. The inset
shows the QMC data
over a larger range of $\beta$ and $cq/T$ together with the
universal scaling function. }
\label{chi}
\end{figure}

\begin{figure}
\caption
{Comparison of QMC data for 
real-space correlation functions ( symbols ) and
Eq.~(\protect\ref{4}) ( solid lines ) with
$T_0=4.5$ and $D=0.075$.
The inset shows the ratio of the equal-time correlation functions
at $\beta=16$ and $32$ compared with Eq.~(\protect\ref{4}) and
the expression with $\Delta=1/4$ \protect\cite{Sachdev}.}
\label{Sr}
\end{figure}

\begin{figure}
\caption
{The ratio $\protect T_{2G}/\protect \sqrt{T} T_1$
versus $T$ from Eqs.~(\protect\ref{9}) and (\protect\ref{10}). 
The $T=0$ limit of the ratio is $1.68$. The inset shows a linear variation 
with $1/\ln{(\beta)}$. QMC and HTE data for the 
ratio with full $T_{2G}$ and with only the scaling part included are 
shown by the symbols.}
\label{ratio}
\end{figure}

\begin{figure}
\caption
{Imaginary part of the antiferromagnetic 
susceptibility. Lines represent Eq.~(\protect\ref{11}) with parameters as
in Fig.~\protect\ref{Sr} and the symbols 
represent the QMC + max-ent data.}
\label{chi''}
\end{figure}


\begin{references}
\bibitem[*]{byline1} Present address: Department of Physics, University of 
Illinois at Urbana-Champaign, Urbana, IL 61801
\bibitem{CHNCSY} S. Chakravarty, B. I. Halperin and D. R. Nelson, \prb
{\bf 39}, 2344 (1989); A. V. Chubukov, S. Sachdev and J. Ye, \prb {\bf 49},
11919 (1994).
\bibitem{Neutron} D. A. Tennant, R. A. Cowley, S. Nagler and A. M. Tsvelik,
\prb {\bf 52}, 13368 (1995).
\bibitem{Takigawa} M. Takigawa, N. Motoyama, H. Eisaki and S. Uchida,
\prl {\bf 76}, 4612 (1996).
\bibitem{Schulz} H. J. Schulz, \prb {\bf 34}, 6372 (1986).
\bibitem{Shankar} R. Shankar, Int. J. Mod. Phys. B {\bf 4}, 2371 (1990).
\bibitem{SSS} S. Sachdev, T. Senthil, and R. Shankar, \prb {\bf 50}, 258 
(1994).
\bibitem{Giam} T. Giamarchi and H. J. Schulz, \prb {\bf 39}, 4620 (1989);
R. R. P. Singh, M. E. Fisher and R. Shankar, \prb {\bf 39}, 2562 (1989).
\bibitem{2-spinon} A. H. Bourgourzi, M. Couture, and M. Kacir,
q-alg/ 9604019; M. Karbach, G. Muller, and A. H. Bourgourzi,
cond-mat/9606068.
\bibitem{KM} T. Koma and N. Mizukoshi, J. Stat. Phys. {\bf 83}, 661 (1996)
\bibitem{Cardy} J. L. Cardy, J. Phys. A {\bf 17}, L385 (1984)
\bibitem{AB} I. Affleck and J. C. Bonner, \prb {\bf 42}, 954 (1990);
I. Affleck, D. Gepner, H. J. Schulz, and T. Ziman,
J. Phys. A {\bf 22}, 511 (1989)
\bibitem{Nomura} K. Nomura and M. Yamada, \prb {\bf 43}, 8217 (1991)
\bibitem{Sandvik1}A. W. Sandvik and J. Kurkij\"arvi, \prb {\bf 43}, 
5950 (1991);
A. W. Sandvik, J. Phys. A {\bf 25}, 3667 (1992).
\bibitem{Sachdev} S. Sachdev, \prb {\bf 50}, 13006 (1994)
\bibitem{Eggert} S. Eggert, I. Affleck, and M. Takahashi, \prl {\bf 73},
332 (1994).
\bibitem{Starykh} O. A. Starykh, A. W. Sandvik and R. R. P. Singh,
to be published.
\bibitem{maxent} M. Jarrell and J. E. Gubernatis, Phys. Rep.
{\bf 269}, 133 (1996).
\bibitem{footnote} The numerical data are presented with $R=-0.5$ 
\cite{Sandvik}.
\bibitem{Sandvik}A. W. Sandvik, \prb {\bf 52}, R9831 (1995).
\bibitem{Neutron2} D. C. Dender, D. H. Reich, C. Broholm, K. Lefmann
and G. Aeppli, F9-3, Bull. Amer. Phys. Soc. 
March 1996 .
\bibitem{Eggert2} S. Eggert, cond-mat/9602026.
\end{references}
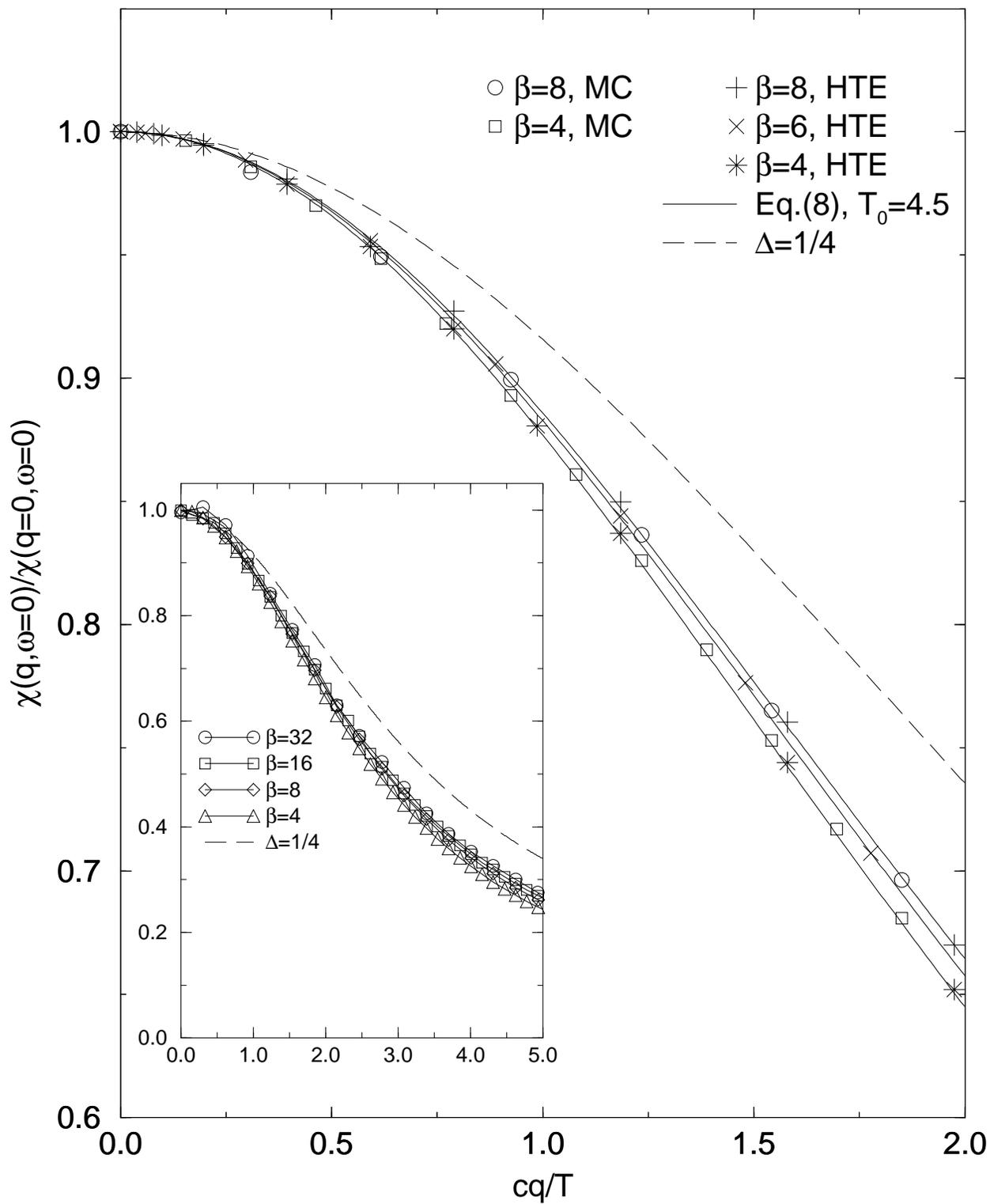
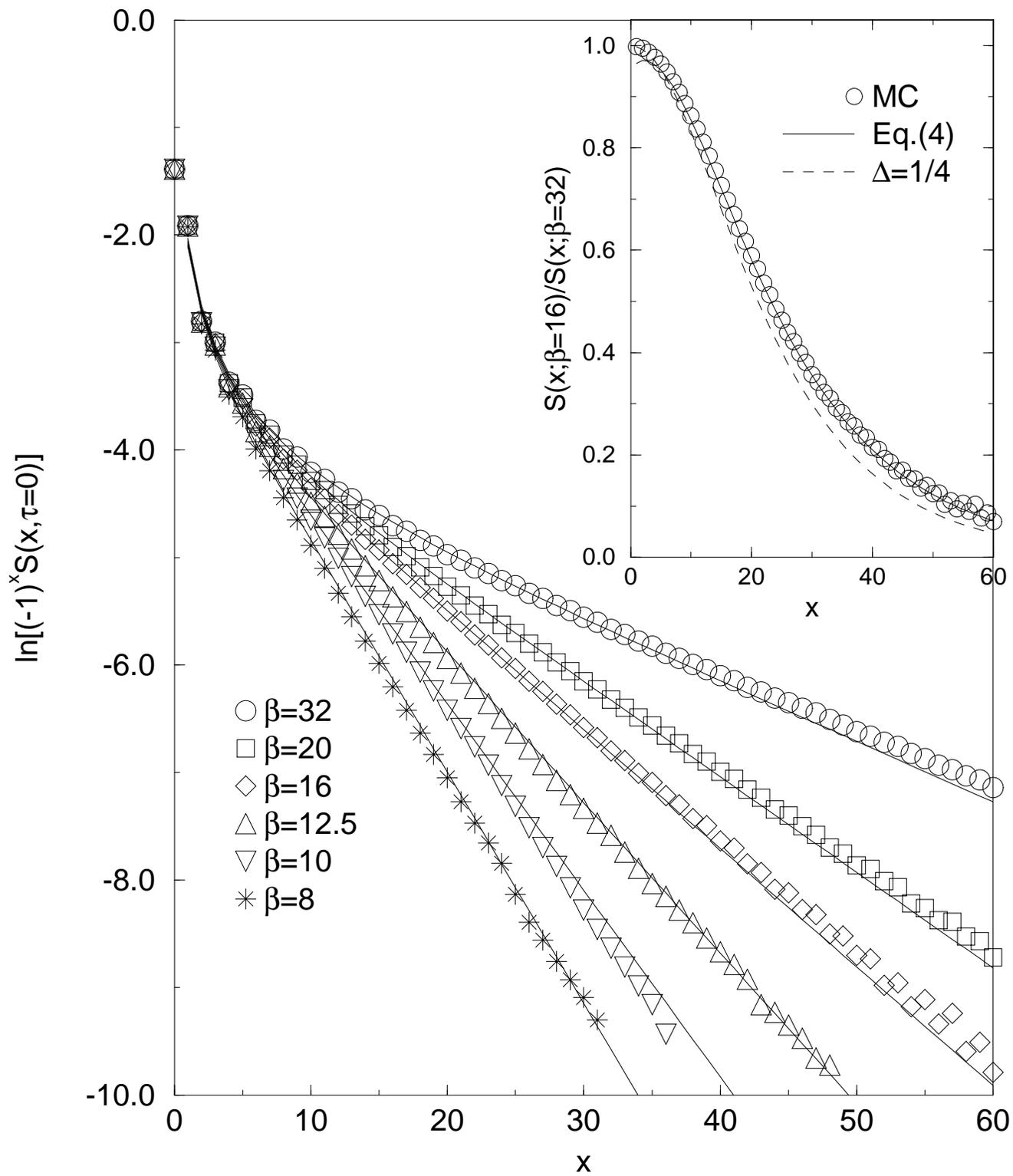
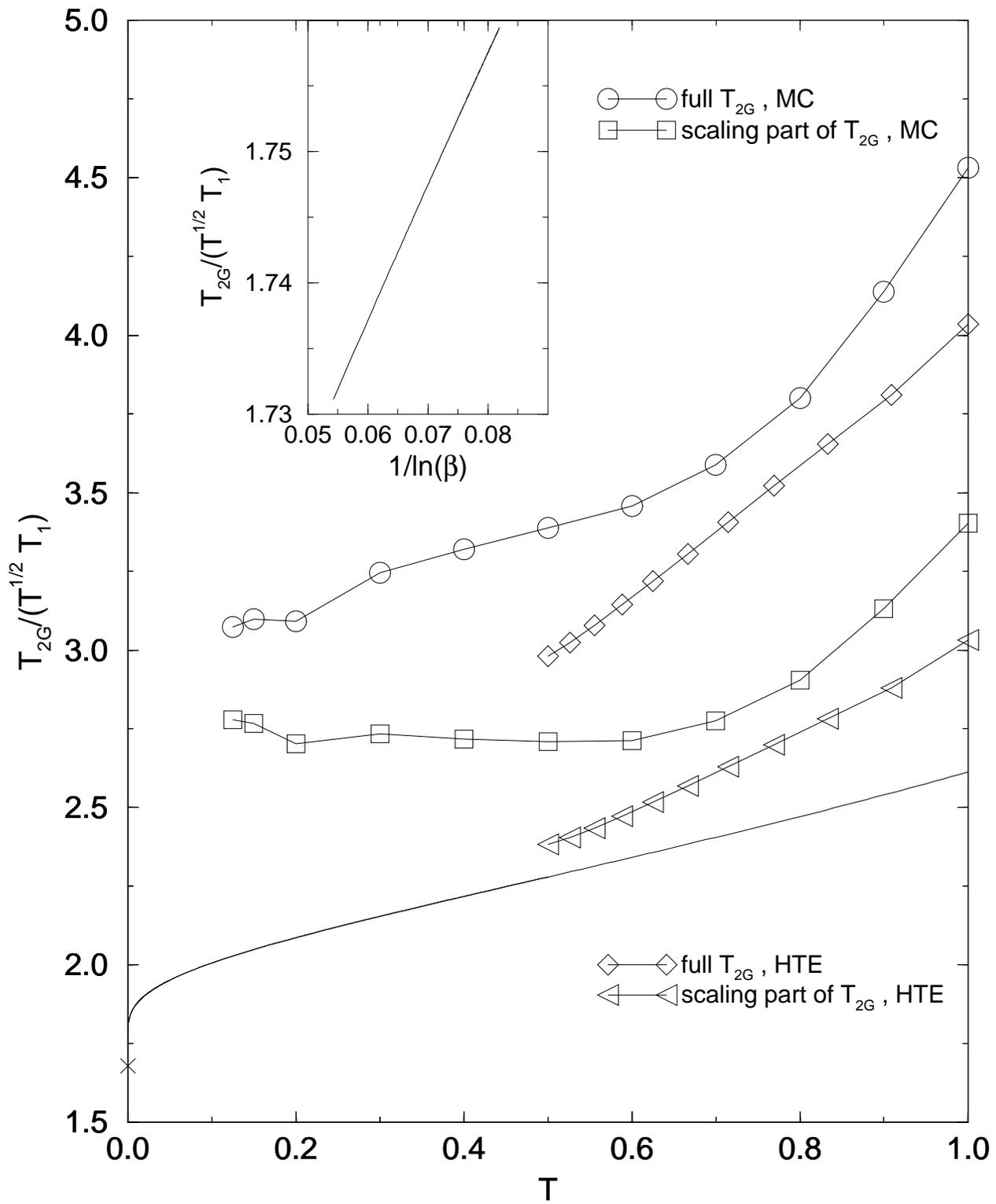
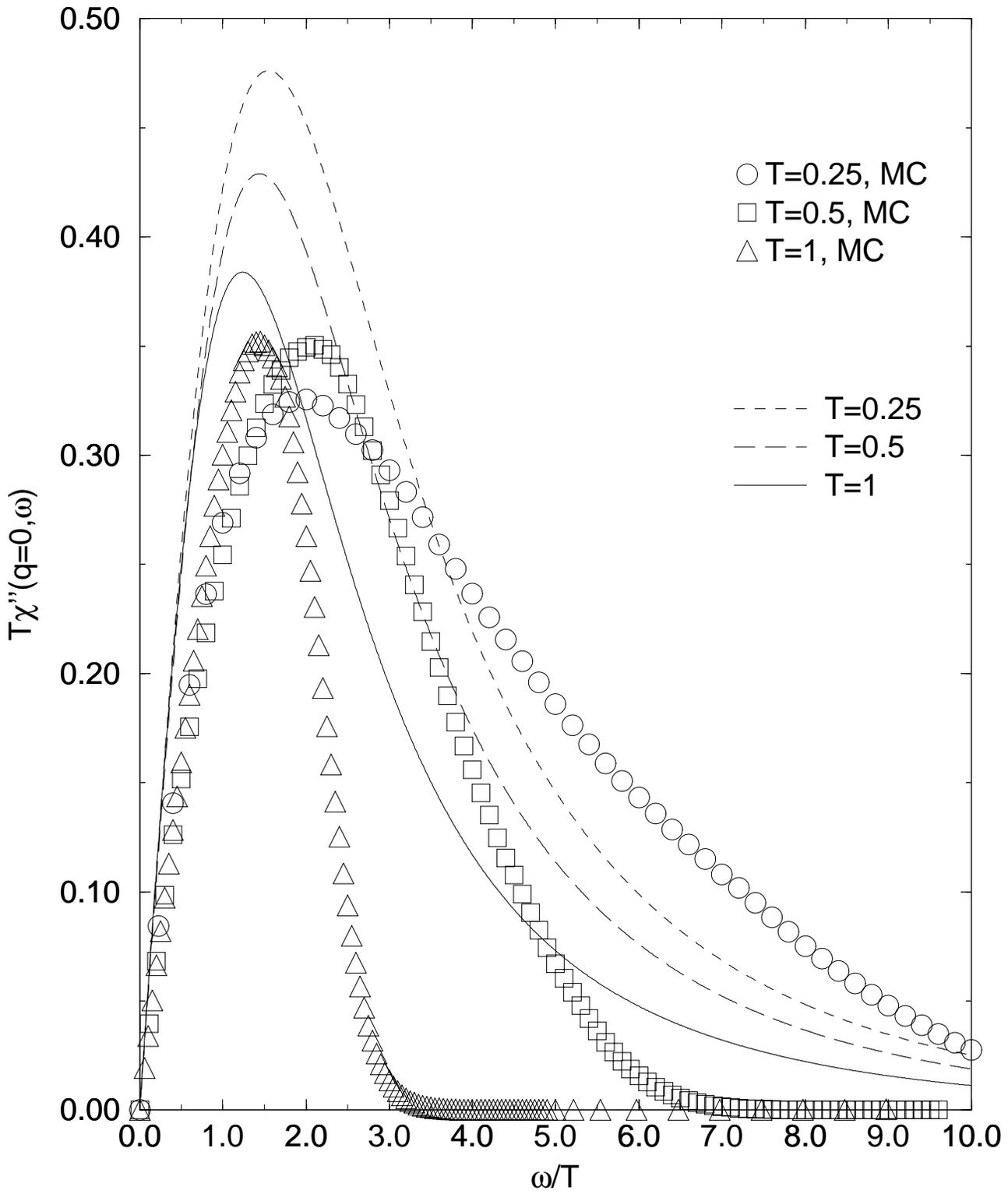
\end{document}